\documentclass[%
 reprint,
 amsmath,
 amssymb,
 aps,
 prl,
]{revtex4-2}

\usepackage{graphicx}
\usepackage{dcolumn}
\usepackage{bm}
\usepackage{xcolor}
\usepackage{hugopaper} 

\begin{document}

\title{Salt crystallization and deliquescence triggered by humidity cycles in nanopores}

\author{Hugo Bellezza}
\author{Marine Poizat}
\author{Olivier Vincent}
 \email{olivier.vincent@cnrs.fr}
\affiliation{
    Universite Claude Bernard Lyon 1, CNRS, Institut Lumi\`ere Mati\`ere,
	UMR5306, F-69100, Villeurbanne, France
}

\date{\today}

\begin{abstract}

    We study the response of materials with nanoscale pores containing sodium chloride solutions, to cycles of relative humidity (RH).
    Compared to pure fluids, we show that these sorption isotherms display much wider hysteresis, with a shape determined by salt crystallization and deliquescence rather than capillary condensation and Kelvin evaporation.
    Both deliquescence and crystallization are significantly shifted compared to the bulk and occur at unusually low RH.
    We systematically analyze the effect of pore size and salt amount, and rationalize our findings using confined thermodynamics, osmotic effects and classical nucleation theory.

\end{abstract}

\maketitle


Salt water is abundantly present on Earth.
When lifted into the air as sea spray, it forms microdroplets, whose evolution depend on local temperature and air relative humidity (RH).
In sufficiently dry conditions, evaporation leads to salt crystallization, whereas increasing RH may induce spontaneous crystal dissolution from water vapor molecules, i.e., deliquescence \cite{Davis2015a,Tang1997}.
Crystallization and deliquescence are thus important processes dictating the fate of aerosols, formation of clouds and radiative forcing in the atmosphere, which currently constitue the largest unknowns in climate models \cite{Gryspeerdt2023,Virtanen2025}.

Salt water also interacts with terrestrial and built environments: in coastal and desert landscapes, deposition of salt increases cohesion and modifies dune dynamics or dust emission \cite{Li2021}; for agriculture, it poses an increasing problem due to soil salinization \cite{Hassani2021}; in urban settings, it can penetrate buildings and artworks through sea spray deposition or winter road salting, and persists as brine in geophysical layers.
In such confined situations, crystallization and deliquescence cycles can cause material damage, posing challenges for civil engineering, heritage preservation or underground gas storage \cite{Scherer2004,Rijniers2005,Schiro2012,Osselin2015}.
In technology, salt crystallization is detrimental for desalination or gas-separation membranes \cite{Olufade2018,Engarnevis2020}, but crystallization/deliquescence cycles can also be exploited for energy conversion, heat storage or water harvesting \cite{Posern2015,Garzon-Tovar2017,Kallenberger2018a}.
In all those contexts, understanding the conditions that trigger phase change is crucial in order to avoid or promote precipitation/dissolution, predict when and where the transitions will occur, and understand their impact.

While there has been significant progress in understanding spatial patterns of salt crystallization \cite{Veran-Tissoires2012,Shahidzadeh2015,Mailleur2018,Kohler2018}, crystal growth dynamics and related stresses \cite{Desarnaud2016,Naillon2018,Kohler2022}, or supersaturations achievable before nucleation \cite{Desarnaud2014,Naillon2015,Cedeno2023}, most studies considered millimeter down do micrometer-sized spaces (pores, capillaries, microfluidic channels).
In contrast, little is currently known about if and how nanoscale confinement scale modify these phenomena.

\begin{figure}
    \includegraphics[scale=1]{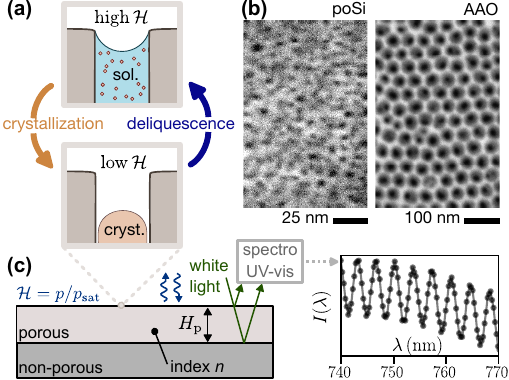}
    \caption{
        \label{fig:Intro}
        (a) Cycles of relative humidity ($\rh = p / \psat$) trigger crystallization and deliquescence of salt solutions upon $\rh$ decrease or increase, respectively.
        (b) We study these phase transitions in mesoporous samples made of silica (poSi) or alumina (AAO); the micrographs are scanning electron microscope observations.
        (c) With white light interferometry (WLI), we extract from periodic modulations in reflectance spectra the optical path length ($\opl = 2 n \idx{H}{p}$), which varies as a function of water content in the pores.
    }
\end{figure}

On the other hand, there exists a vast literature on phase change of pure fluids in nanoscale confinement due to numerous applications including catalysis, biomaterials, sensors, actuators etc. \cite{Huber2015}.
In particular, mesoporous materials (pore diameters \qtyrange{2}{50}{\nm}) display a hysteretic response to vapor pressure cycles (sorption isotherms), often explained with shifted, equilibrium evaporation and metastable condensation \cite{Monson2012,Thommes2014}, although recent results are challenging this idea \cite{Haidar2024}.
This simple picture is also questioned when considering disorder, percolation, pore blocking and/or nucleation \cite{Aubry2014,Doebele2020,Sollner2024}.
However, much less is known on how isotherms change when the fluid contains solutes.

Recent studies with sodium chloride (NaCl) in mesoporous silica have suggested sterically hindered crystallization \cite{Jain2019}, or shifted deliquescence due to curvature-induced modifications of thermodynamic constants \cite{Talreja-Muthreja2022}.
However, a consistent framework describing simultaneously humidity-induced crystallization and deliquescence, and more generally predicting the hysteretic behavior of nanoporous systems containing salts, is still lacking.
Here, we report a systematic investigation of the response of a variety of mesoporous materials to RH cycling [Fig.\ \ref{fig:Intro}(a)], as a function of both pore size (down to diameters $< \qty{4}{\nm}$) and amount of salt.
We show that RH hysteresis is controlled by deliquescence and crystallization rather than capillary condensation and evaporation for pure fluids, thus reversing the equilibrium and metastable branches of the sorption isotherms.
We also exclude steric hindrance as the crystallization mechanism as was supposed earlier \cite{Jain2019}, and demonstrate kinetically limited nucleation at extremely large supersaturations and record low RH.
We rationalize the observed dependencies with modeling based on confined thermodynamics, osmotic effects, and modified classical nucleation theory (CNT).


We used 9 mesoporous samples made of silica (oxidized porous silicon, \emph{poSi}) or anodic aluminum oxide (\emph{AAO}), with average pore diameter spanning $< \qty{4}{\nm}$ to \qty{20}{\nm}, see Fig.\ \ref{fig:Intro}(b) and Supplemental Material \footnote{See \emph{Supplemental Material} for details on the experimental setup; sample fabrication, characterization and preparation; discussion on the impact of contact angle in the theory; and calculations of properties of water and salt solutions.
}.
Since samples consisted of thin ($\idx{H}{p}=5$ to 75 \unit{\um}), transparent porous layers with parallel faces, we used white light interferometry (WLI) to track their water content.
WLI consists in analyzing reflectance spectra, $I(\lambda)$, which present modulations due to light interference between rays reflected on top and at the bottom of the layer [Fig.\ \ref{fig:Intro}(c)].
Spectra are periodic in wavenumber $k = 2 \pi / \lambda$, so that Fourier transform of $I(k)$ directly yields the optical path length, $\opl = 2 n \idx{H}{p}$, where $n$ is the refractive index \cite{Casanova2012}.
Evaporation and condensation of water modify $n$, resulting in variations of $\opl$ that are approximately proportional to the pore filling fraction \cite{Bossert2020}.

\begin{figure}
    \includegraphics[scale=1]{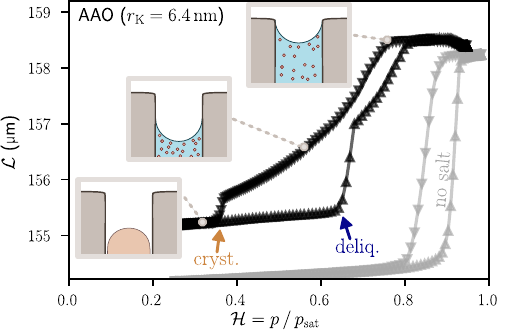}
    \caption{
        \label{fig:IsothWaterSalt}
        Water sorption isotherms measured with WLI on an AAO sample with (black, $\idx{m}{i} = \qty{3.2}{\mol\per\kg}$) and without (grey, $\idx{m}{i} = 0$) salt in the pores.
        The direction of the symbols indicate increasing ($\triangle$) or decreasing ($\triangledown$) $\rh$.
        Sketches represent the inferred status of the pore space with the same color coding as in Fig.\ \ref{fig:Intro}, and the arrows point to the crystallization and deliquescence transitions.
    }
\end{figure}

For imposing RH, we used a custom-made vacuum chamber with precise control and measurement of the (absolute) vapor pressure of water, $p$, and directly calculated the corresponding RH, $\rh = p / \psat(T)$ where $\psat$ is the saturation pressure of water at the temperature of the experiments ($T = \qty{25}{\degreeCelsius}$).
Using this system, we measured water sorption isotherms, which consist in tracking variations of water content (i.e., $\opl$) as a function of slowly increasing or decreasing RH ($a = \odv{\rh}/{t} \simeq \qtyrange{e-4}{e-3}{\per\second}$, i.e., typical cycle times of $2 / a \simeq \numrange{0.5}{5}$ hours).
We repeated each isotherm measurement several times to check reproducibility.
Across all samples and experimental conditions, we measured more than 500 individual isotherms.

We first recorded isotherms without salt in the pores (Fig.\ \ref{fig:IsothWaterSalt}, gray), from which we estimated the capillary pressure $\DeltaPc$ (\unit{\pascal}) at the evaporation point (relative humidity, $\idx{\rh}{c}$), using the Kelvin-Laplace equation, $\DeltaPc = (\kb T / \vw) \ln \idx{\rh}{c} < 0$, where $\idx{v}{w}$ is the molecular volume of liquid water (\unit{\meter\cubed}), and $\kb T$ thermal energy (\unit{\joule}) \cite{Note1}.
We also define the sample-dependent \emph{Kelvin radius}, $\rk = - 2 \sigmaw / \DeltaPc$, where $\sigmaw$ is the water surface tension (\unit{\newton\per\meter}),
which corresponds to the radius of curvature of the liquid-vapor menisci in the pores at the evaporation point, and is related to the geometrical pore radius, $\rp$, through $\rk = \rp / \cos \theta$, with $\theta$ the receding contact angle of the fluid on the pore walls \cite{Note1,Factorovich2014a}.
Analysis on representative samples shows that $\rk$ matches closely the geometrical pore radius, $\rp$, estimated from nitrogen sorption isotherms, suggesting that $\rk$ is a good in-situ estimate of pore dimensions and that $\theta$ is close to 0 for all studied samples \cite{Note1}.
As we will see below, $\rk$ is also a more natural quantity than $\rp$ when predicting the RH of crystallization and deliquescence.

Then, we introduced \ce{NaCl} in the pore space by spontaneous imbibition of salt solutions \cite{Note1}.
Since \ce{NaCl} is nonvolatile, it remained trapped in the sample for the duration of the experiments.
Salt could be later removed by rinsing in pure water, allowing us to repeat this procedure to vary the in-pore salt concentration (molality, $\idx{m}{i}$, \unit{\mol\per\kg}).
We extracted $\idx{m}{i}$ from WLI, taking advantage of the loss of contrast in the interference signal when $\rh$ exceeds the equilibrium RH of the solution \cite{Note1}.
Our experimental setup and procedure thus allowed for in-situ measurement of both pore dimensions ($\rk$) and amount of salt ($\idx{m}{i}$).
In the following, we systematically evaluate the influence of these two parameters on isotherms.


Fig.\ \ref{fig:IsothWaterSalt} (black curve) shows a typical isotherm obtained on an AAO sample ($\rk = \qty{6.4(1.6)}{\nm}$), with salt ($\idx{m}{i} = \qty{3.2}{\mole\per\kg}$).
Compared to the case without salt (gray curve), several qualitative changes are visible.
First, salt globally shifts the curves to lower RH, which is expected due to colligative solute-induced vapor pressure decrease \cite{Vincent2019,Jain2019}.
Second, the hysteresis between condensation and evaporation is much wider than with pure water, due in particular to a progressive increase of concentration of the solution by continuous meniscus recession on the evaporation branch (see sketches in Fig.\ \ref{fig:IsothWaterSalt}).
Last, we observe two sharp, almost vertical transitions at $\rh \simeq 0.65$ and $\rh \simeq 0.35$.
Because these transitions correspond to a sudden intake or loss of liquid from the pores and separate a situation with solution in the pores (high $\rh$) and one with crystal (low $\rh$), we attribute them to deliquescence and crystallization, respectively (arrows in Fig.\ \ref{fig:IsothWaterSalt}).

\begin{figure}
    \includegraphics[scale=1]{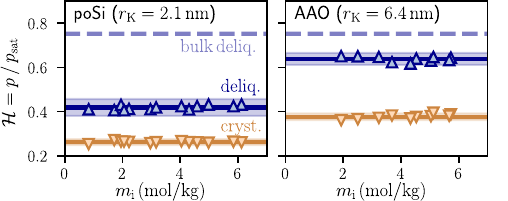}
    \caption{
        \label{fig:ConcentrationEffect}
        Effect of initial concentration ($\idx{m}{i}$) on crystallization and deliquescence RH for two samples with different pore sizes.
        Uncertainties are on the order of symbol sizes or smaller.
        The dashed blue line corresponds to bulk crystal-solution equilibrium for NaCl ($\rh_0 = 0.753$).
        Continuous brown and blue lines correspond to data average for crystallization and deliquescence respectively, surrounded by shaded areas which represent the scatter of the data.
        }
\end{figure}

We tested typically $\simeq 10$ concentrations in the range $0 < \idx{m}{i} \leq m_0$ in every sample, where $m_0 = \qty{6.15}{\mol\per\kg}$ is NaCl solubility in water \cite{Note1}.
For each experiment, we estimated the position of the onsets deliquescence and crystallization, extracted from the range where the isotherm data such as in Fig. \ref{fig:IsothWaterSalt} exhibited maximum curvature.
For some poSi samples (and only at large $\idx{m}{i}$), we observed two distinct crystallization steps.
In such situations, we considered only the final one (lowest $\rh$), which was always consistent with the single crystallization steps observed with the same samples at lower salt amount.

Fig.\ \ref{fig:ConcentrationEffect} shows results obtained with one AAO sample (left) and one poSi sample (right).
These results indicate obvious confinement effects, since deliquescence occurs at significantly lower RH than in bulk ($\rh_0 = 0.753$ \cite{Note1}, dashed blue line).
Also, Fig.\ \ref{fig:ConcentrationEffect} shows that RH for both transitions are approximately independent of the amount of salt.

While this independence may seem surprising, it can be explained by examining the equilibrium between pore water and external vapor, which imposes $\Psi (\rh) = \DeltaPc (\rk) - \Pi(m)$ where $\Psi = \kb T \ln \rh / \vw$ is the \emph{water potential} of the vapor and $\Pi(m)$ the osmotic pressure of the solution \cite{Vincent2019}.
We consider situations where the meniscus has receded inside the pore (such as in the central sketch in Fig.\ \ref{fig:IsothWaterSalt}) and reached a curvature imposed by pore geometry and contact angle.
We also assume that the corresponding capillary pressure, $\DeltaPc$, does not significantly change with salt concentration, because increases in surface tension with concentration are compensated by an increase of contact angle \cite{Sghaier2006,Zhang2015b}.
As a result, $\DeltaPc = - 2 \sigmaw / \rk$ for all concentrations, with $\sigmaw$ the surface tension of pure water.
Finally, osmotic pressure is related to solute concentration through  $\Pi = \nu \phi \kb T \idx{\rho}{w} m $, with $\nu = 2$ the salt dissociation number, $\idx{\rho}{w}$ pure water density (\unit{\kg\per\m\cubed}) and $\phi(m)$ the osmotic coefficient, which accounts for deviations to ideality and tends toward 1 at low concentrations \cite{Note1}.
As a result, the liquid/vapor equilibrium imposes
\begin{equation}
    \label{eq:LiquidVaporEquilibrium}
    - \ln \rh = \frac{\idx{r}{\ell}}{\rk} + \alpha_0 \, \phi  \, S(m)
\end{equation}
with $S = m / m_0$ the supersaturation, $\idx{r}{\ell} = 2 \sigmaw \vw / (\kb T)$ a typical length that depends only on pure liquid properties and temperature ($\idx{r}{\ell} \simeq \qty{1.05}{\nm}$ for water at \qty{25}{\degreeCelsius}) and $\alpha_0 = \idx{M}{w} m_0$ ($\idx{M}{w}$ molar mass of water).

\begin{figure}
    \includegraphics[scale=1]{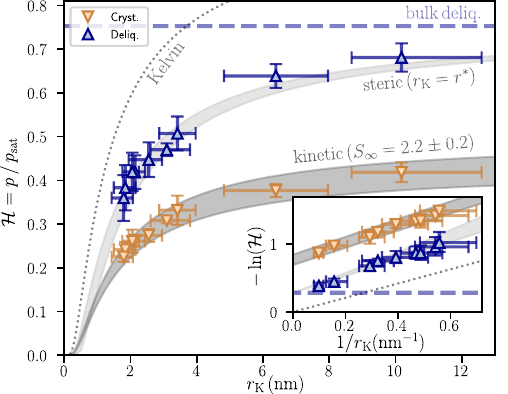}
    \caption{
        \label{fig:CrystDeliq_All}
        Crystallization and deliquescence RH as a function of pore size.
        Triangle symbols represent experimental data.
        Uncertainties in $\rh$ correspond to the observed scatter across experiments at different $\idx{m}{i}$ (shaded areas in Fig.\ \ref{fig:ConcentrationEffect}).
        Uncertainties in $\rk$ correspond to the width of the pore size distribution as estimated from sorption isotherms \cite{Note1}.
        The dashed blue line shows bulk deliquescence.
        The dotted black line represents the confined equilibrium of pure water as predicted by the Kelvin-Laplace equation.
        The light gray area corresponds to our theoretical prediction for sterically limited nucleation; the extension of this zone originates from the uncertainty in $\sigmacl$.
        The dark gray area corresponds to kinetically limited nucleation with a bulk supersaturation, $S_\infty = 2.2 \pm 0.2$.
        The inset displays the same data with transformed axes reflecting the relation predicted by Eq.\ (\ref{eq:LiquidVaporEquilibrium}).
        }
\end{figure}

Eq.\ (\ref{eq:LiquidVaporEquilibrium}) indicates that for a given sample of radius $\rk$, there is a direct relation between imposed RH and concentration, $m$, irrespective of the initial concentration $\idx{m}{i}$; if the sample has less salt initially (small $\idx{m}{i}$), more water should evaporate at a given RH, in order to achieve the same value of $m$.
This effect explains the independence of the deliquescence and crystallization transitions as a function of $\idx{m}{i}$ observed in Fig.\ \ref{fig:ConcentrationEffect}, and suggests that these transitions occur at a well-defined concentration.
Because of this independence, it is also possible to assign for each sample (characterized by $\rk$) a unique value for the deliquescence and crystallization RH, that we have reported in Fig.\ \ref{fig:CrystDeliq_All} for all 9 studied samples.

Fig.\ \ref{fig:CrystDeliq_All} shows that the observed transitions occur at RH far below both bulk deliquescence (dashed blue line) and pure liquid evaporation predicted by the Kelvin-Laplace equation (dotted black line).
The data also displays a clear trend where both crystallization and deliquescence shift to lower RH when the pore size is reduced, as reported recently for deliquescence in larger pores \cite{Talreja-Muthreja2022}.
Remarkably, the crystallization RH are all below the lowest values reported in the literature for \ce{NaCl} solutions ($\rh \simeq \qtyrange{40}{45}{\percent}$RH, see e.g. Refs \cite{Chan1997,Gao2007}).

Inspired by Eq.\ (\ref{eq:LiquidVaporEquilibrium}) we also replotted the data as $-\ln \rh$ as a function of $1 / \rk$ (Fig.\ \ref{fig:CrystDeliq_All}, inset).
In this representation, the x-axis represents confinement, and shifts along the y-axis indicate increased degrees of supersaturation, so that from the location of the data we anticipate that the concentration at crystallization is much larger than during deliquescence.

\begin{figure}
    \includegraphics[scale=1]{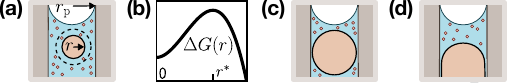}
    \caption{
        \label{fig:CNT}
        (a) In-pore growth of a spherical nucleus,
        (b) Corresponding Gibbs free energy profile from CNT (Eq. \ref{eq:CNT_G_dimensionless}).
        (c) Situation where $r=\rp=r^\ast$.
        (d) Three-phase equilibrium described in Ref.\ \cite{Talreja-Muthreja2022}.
    }
\end{figure}

In order to quantitatively predict the observed shifts, we first test the hypothesis that crystallization occurs when the critical radius of nucleation, $r^\ast$, becomes lower than the pore radius $\rp \simeq \rk$, i.e. due to steric hindrance as proposed in Ref.\ \cite{Jain2019}.
Following a modified classical nucleation theory (CNT),
\begin{equation}
    \label{eq:CNTCriticalRadius}
    r^\ast = \frac{
        2 \sigmacl \vc
    }{
        \nu \kb T \ln (\gamma S)
        +
        \DeltaPc (\rk) \Delta v
    }
\end{equation}
where $\sigmacl$ is the surface energy of the crystal-solution interface (\unit{\newton\per\meter}), $\vc$ is the molecular volume of the crystal (\unit{\meter\cubed}), $\gamma = \gamma_\pm / (\gamma_\pm)_0$ is a ratio of \ce{NaCl} mean activity coefficients with saturation as a reference, and $\Delta v = \vs - \vc$ where $\vs$ is the partial molecular volume of the solute in solution, which can be calculated from solution density \cite{Note1,Jain2019}.
Eq.\ (\ref{eq:CNTCriticalRadius}) assumes homogeneous nucleation, i.e., that the pore wall does not promote crystallization and that spherical nuclei appear in the pore liquid without contact with the walls, see Fig.\ \ref{fig:CNT}(a).
The most uncertain parameter in Eq.\ (\ref{eq:CNTCriticalRadius}) is $\sigmacl$, with a variety of values in the range $\qtyrange{30}{90}{\milli\newton\per\meter}$ reported from various experimental techniques including fitting of nucleation data \cite{Peckhaus2016,Talreja-Muthreja2022}, up to more than \qty{100}{\milli\newton\per\meter} from molecular dynamics simulations \cite{Sanchez-Burgos2023}.
We choose $\sigmacl = \qty{70(40)}{\milli\newton\per\meter}$ to represent this large range in a conservative manner.

The effect of confinement is represented by the second term of the denominator in Eq.\ (\ref{eq:CNTCriticalRadius}), which accounts for the negative capillary pressure due to the curvature of the liquid-vapor interface.
Since $\Delta v < 0$ for \ce{NaCl} in water and $\DeltaPc = - 2 \sigmaw / \rk < 0$, this effect reduces the critical radius $r^\ast$, i.e., facilitates nucleation in confinement.

In order to describe sterically hindered nucleation, we set $r^\ast = \rk$ (because $\rp \simeq \rk$) in Eq.\ (\ref{eq:CNTCriticalRadius}), leading to an implicit relation between $S = m / m_0$ and $\rk$ that we solve numerically, since $\gamma$ and $\Delta v$ depend on $m$ in non trivial manners \cite{Note1}.
We then use Eq.\ (\ref{eq:LiquidVaporEquilibrium}) to find from $S(\rk)$ the corresponding RH (light gray shaded area in Fig.\ \ref{fig:CrystDeliq_All}).
Remarkably, the model does not have fitting parameters, and the impact of the poorly known parameter $\sigmacl$ is small (see extension of light gray zone).
While this prediction obviously fails to account for the experimental crystallization data (brown, downward triangles), it surprisingly fits very well the deliquescence data (blue, upward triangles).

If this coincidence seems unexpected, Eq.\ (\ref{eq:CNTCriticalRadius}) with $r^\ast = \rk$ is in fact identical to recent thermodynamic predictions for the three-phase equilibrium between water vapor, solution and salt crystal in confinement, assimilating $\rk$ to $\rp$ and neglecting terms related to crystal compressibility and adsorbed films (Eq.\ (10) in Ref.\ \cite{Talreja-Muthreja2022}).
We explain this convergence by the fact that $r=r^\ast$ represents a maximum of Gibbs free energy in CNT [Fig.\ \ref{fig:CNT}(b)], thus $\rp = r^\ast$ describes the unstable equilibrium of a crystal with radius of curvature, $\rp$ [Fig.\ \ref{fig:CNT}(c)].
On the other hand, Ref.\ \cite{Talreja-Muthreja2022} also examines an equilibrium situation where the crystal-solution interface has a curvature determined by the pore radius [Fig.\ \ref{fig:CNT}(d)], so that the equilibrium conditions corresponding to Fig.\ \ref{fig:CNT}(c) and Fig.\ \ref{fig:CNT}(d) are identical.
Thus, our theory and the equations from Ref.\ \cite{Talreja-Muthreja2022} correspond to an unstable equilibrium which equally describes spontaneous dissolution and sterically hindered crystal nucleation.
This remark indicates that if crystallization was limited by steric hindrance, crystallization and deliquescence should occur at the same RH, which is clearly not the case (see Fig. \ref{fig:CrystDeliq_All}).

If crystal nucleation is not sterically limited, it has to be determined by kinetics, i.e., thermal fluctuations.
CNT estimates nucleation rates from the energy barrier as in Fig.\ \ref{fig:CNT}b, however with uncertainties related to unknowns on $\sigmacl$ and on kinetic prefactors \cite{Cedeno2023}.
To avoid committing to specific values for these parameters, we normalize our confined CNT model to bulk CNT, resulting in
\begin{equation}
    \label{eq:CNT_Supersat}
    \gamma S = \gamma_\infty S_\infty \exp \left( - \frac{\DeltaPc(\rk) \Delta v}{\nu \kb T}  \right)
\end{equation}
for the supersaturation required for kinetic nucleation, where $S_\infty$ and $\gamma_\infty$ correspond to nucleation in bulk under the same conditions, see \emph{End Matter} for details.

We numerically solved Eq.\ (\ref{eq:CNT_Supersat}) to find $S(\rk)$, using $S_\infty$ as a fitting parameter; we then calculated the corresponding crystallization RH with Eq.\ (\ref{eq:LiquidVaporEquilibrium}).
We found that we could account for our experimental data using $S_\infty = \num{2.2(0.2)}$ (dark gray shaded area in Fig.\ \ref{fig:CrystDeliq_All}).
This value matches the largest reported supersaturations for \ce{NaCl}, usually assumed to correspond to homogeneous nucleation, which were obtained with oil-immersed microdroplets \cite{Cedeno2023} or levitated droplets \cite{Cohen1987a,Na1994,Gao2007} and lie in the range $S=\num{2.15(0.25)}$ \footnote{Ref.\ \cite{Chan1997} also lists activity data down to $m=\qty{16.68}{\mole\per\kg}$, i.e., $S=2.7$ but without indications on repeatability and average values.}.
Thus, despite the very large area of contact between the liquid and pore walls, NaCl nucleation apparently occurs homogeneously, suggesting that crystal-wall interactions are not favorable.

Note that Eqs.\ (\ref{eq:LiquidVaporEquilibrium}) and (\ref{eq:CNT_Supersat}) depend directly on $\rk$, which makes Kelvin radius the relevant pore dimension to describe confined crystal nucleation, rather than $\rp$ (see also Supplemental Material \cite{Note1}).
In fact, even for deliquescence, differences between $\rk$ and $\rp$, e.g. due to nonzero contact angle ($\theta$), result in negligible differences in the predictions of the transition RH, even for $\theta$ as high as $\qty{45}{\degree}$ \cite{Note1}.
Finally, theory lines in Fig. \ref{fig:CrystDeliq_All} predict a crossover of the steric and kinetic limits at $\rk < \qty{1}{\nm}$, implying that NaCl crystallization could become sterically hindered for extremely small pore sizes.
It is however not obvious whether continuum-based approach would still be valid at such extreme confinements comparable to molecular dimensions.


In conclusion, we have shown that in nanoporous materials containing NaCl solutions, humidity cycling triggers crystallization and deliquescence at record-low RH far below bulk values, and that the confined solution can sustain extremely large supersaturations.
By systematically analyzing the role of pore size and salt content, we have demonstrated unambiguously that crystallization is not limited by steric hindrance as previously assumed, but occurs through kinetic nucleation in the metastable solution, while deliquescence corresponds to a sterically-defined (unstable) equilibrium.
These transitions fundamentally change the mechanisms responsible for the shape and hysteresis of sorption isotherms, which involve equilibrium evaporation and (metastable) capillary condensation for pure fluids.
This original RH response could inspire new methods to characterize the architecture of nanoporous materials.

Beyond providing a new framework for confined thermodynamics, our findings open perspectives for interpreting phenomena ranging from salt weathering and membrane fouling to humidity responsive materials.
Possibly, our results could also apply to composite aerosols combining solid pores and salt.
Generally, materials with smaller pores uptake water more easily, but are also able to sustain lower humidities without crystallizing.
Salt content does not directly dictate the critical RH of water uptake or release, but defines the amount of water present in the material at a given RH.
Finally, because of large achievable supersaturations, hydrophilic mesoporous materials could also constitute convenient platforms for the fundamental study of deeply metastable solutions.

Future directions include exploring how to further tune the RH response with smaller pores, different surface chemistry and wettability, or salts of different nature.
Also, since our theoretical approach is based on single-pore modeling, there remain outstanding questions about the effect of disorder (pore size distributions, connectivity) and of transport in extended systems.

\paragraph{Acknowledgments ---}

\begin{acknowledgments}

    We thank Mark Busch and Patrick Huber for providing poSi2 samples;
    Panayotis Spathis, Pierre-Etienne Wolf and Laurent Cagnon for providing the AAO2 sample;
    R\'emy Fulcrand for SEM measurements;
    Gilles Simon and Olivier Boisron for help with the vacuum system;
    Sylvie Le Floch for assistance with sample oxidation;
    and L\'eo Martin for early experimental tests.
    We acknowledge discussions with Pierre-Etienne Wolf and Etienne Rolley on WLI, and exchange with Michael Steiger about the activity of salt solutions.
    This work was supported by the French National Research Agency (ANR) under grant ANR-19-CE09-0010 (SINCS).

\end{acknowledgments}

\bibliography{References}

\appendix

\section{End Matter}

Here, we provide a detailed derivation of classical nucleation theory (CNT) for homogeneous nucleation of salt crystals in confinement, leading to Eqs. \ref{eq:CNTCriticalRadius} and \ref{eq:CNT_Supersat}.

\paragraph{CNT: general formulation}

Assuming a spherical nucleus of radius $r$ (Figure \ref{fig:CNT}a), CNT predicts that the Gibbs free energy of nucleation follows
\begin{equation}
    \label{eq:CNT_G_dimensionless}
    \Delta G = \Delta G^\ast \left[ 3 \left( \frac{r}{r^\ast} \right)^2 - 2 \left( \frac{r}{r^\ast} \right)^3 \right]
\end{equation}
(see graph in Figure \ref{fig:CNT}b), with
\begin{equation}
    \label{eq:CNT_CriticalRadius_General}
    r^\ast = \frac{2 \sigmacl \vc}{\Delta \mu}
\end{equation}
\begin{equation}
    \label{eq:CNT_EnergyBarrier_General}
    \Delta G^\ast = \frac{4}{3} \pi (r^\ast)^2 \sigmacl
\end{equation}
Both the critical radius $r^\ast$ and the energy barrier $\Delta G^\ast$ depend on the chemical potential difference (per \ce{NaCl} molecule) $\Delta \mu$ between the salt in solution ($\idx{\mu}{s}$) and in the crystal ($\idx{\mu}{c}$), $\Delta \mu = \idx{\mu}{s} - \idx{\mu}{c}$ \cite{Meldrum2020}.

The nucleation energy barrier $\Delta G^\ast$, never vanishes for finite values of $\Delta \mu$.
Thus, nucleation occurs when this barrier is crossed due to thermal fluctuations (\emph{kinetic nucleation}), with a probability per unit time and per unit volume
\begin{equation}
    \label{eq:CNT_AttemptFrequency_General}
    \Gamma = K \exp \left( - \frac{\Delta G^\ast}{\kb T} \right)
\end{equation}
with $K$ a kinetic prefactor.
Because Eq. \ref{eq:CNT_AttemptFrequency_General} has a very strong dependency in $\Delta \mu$, there is a narrow range of values of $\Delta \mu$ around a value $\idx{\Delta \mu}{k}$, where the probability of nucleation leads to observable crystallization \cite{Meldrum2020}.
For a volume $V$ of solution observed for a time $\tau$, $\idx{\Delta \mu}{k}$ can be estimated with $\Gamma(\idx{\Delta \mu}{k}) V \tau = 1$ \footnote{Sometimes $\Gamma V \tau = \ln 2$ is chosen to account for the evolution of the nucleation probability over time knowing that no nucleation has occurred between time 0 and $t$; this choice has a negligible impact on the final result \cite{Caupin2006}.}, which leads to
\begin{equation}
    \label{eq:CNT_DeltaMuKinetics}
    \idx{\Delta \mu}{k} = \mathcal{C} \kb T
\end{equation}
using Eqs. \ref{eq:CNT_CriticalRadius_General}, \ref{eq:CNT_EnergyBarrier_General} and \ref{eq:CNT_AttemptFrequency_General}, where we have defined the constant
\begin{equation}
    \label{eq:CNT_KineticConstant}
    \mathcal{C} =
    \sqrt{
        \frac{16 \pi}{3 \ln (K V \tau)}
        \left(
            \frac{\sigmacl}{\kb T}
        \right)^3
        \vc^2
    }
\end{equation}
which is typically on the order of \numrange{e1}{e2}.
Since $\mathcal{C}$ depends weakly on the volume $V$, observation time $\tau$ and on the choice for the kinetic prefactor, $K$, the value of $\idx{\Delta \mu}{k}$ for observable nucleation is similar in systems that can vary by several orders of magnitude in experimental conditions, and knowing an exact value for $K$ is not crucial \cite{Caupin2006}.

In order to apply the equations above, we must know how $\Delta \mu$ depends on physical parameters in the system, e.g. salt concentration.
Below, we evaluate $\Delta \mu$ in bulk and in a confined pore liquid and discuss implications for nucleation in pores.

\paragraph{CNT: bulk case}
For an unconfined aqueous solution, $\Delta \mu$ is directly related to the supersaturation and to the activity coefficient through
\begin{equation}
    \label{eq:DeltaMuBulk}
    \Delta \mu = \kb T \ln \Omega
\end{equation}
where we have defined a supersaturation activity ratio
\begin{equation}
    \label{eq:SupersatActivity}
    \Omega = \left( \gamma S \right)^\nu
\end{equation}
with $\gamma$ an activity coefficient ratio that describes deviations to ideality and depends on the concentration of the solution \cite{Steiger2005,Desarnaud2014,Note1}.
From Eq. \ref{eq:CNT_DeltaMuKinetics}, kinetic nucleation thus occurs when the supersaturation in the solution is such that $\Omega = \idx{\Omega}{\infty}$ where
\begin{equation}
    \label{eq:CNT_KineticsBulk}
    \ln \idx{\Omega}{\infty} = \mathcal{C}
\end{equation}

\paragraph{CNT: confined nucleation}

Several effect can modify CNT in a confined situation, e.g., heterogeneous nucleation due to the pore walls, or finite volume effects \cite{Meldrum2020}.
Here, similar to other authors in the context of ice nucleation \cite{Fukuta1966,Denoyel2002} or salt crystallization \cite{Jain2019}, we make the simpler assumption that nucleation of crystal in the pores occurs in a homogeneous manner similarly as in bulk.

In such a situation, confinement impacts nucleation in two ways.
First, steric effects prevent nucleation if the pore radius, $\idx{r}{p}$ is smaller than the critical radius for nucleation, $r^\ast$ \cite{Denoyel2002,Jain2019} (steric hindrance).
Second, because of the curvature of the liquid-vapor interface, the capillary pressure in the liquid phase, $\DeltaPc = - 2 \sigmaw / \rk < 0$ impacts both the chemical potential of salt in the solution and that in the crystal phase \cite{Fukuta1966,Jain2019}, resulting in an additional term in the chemical potential difference
\begin{equation}
    \label{eq:DeltaMuConfined}
    \Delta \mu = \kb T \ln \Omega + \DeltaPc \Delta v
\end{equation}
where $\Delta v = \idx{v}{s} - \vc$ with $\idx{v}{s}$ the partial molecular volume of the salt in the solution.
For \ce{NaCl} in water, $\Delta v < 0$ \cite{Note1}.
This chemical potential shift impacts both the critical radius (Eq. \ref{eq:CNT_CriticalRadius_General}) and the kinetic nucleation limit (Eq. \ref{eq:CNT_DeltaMuKinetics}).

For the critical radius, the combination of Eqs. \ref{eq:CNT_CriticalRadius_General}, \ref{eq:SupersatActivity}, \ref{eq:DeltaMuConfined} directly yields Eq. \ref{eq:CNTCriticalRadius} in the Main Text.
Sterically-limited nucleation can occur only if the pore can accommodate the critical radius, i.e., $r^\ast \leq \idx{r}{p}$.
Using Eq. \ref{eq:CNTCriticalRadius}, this condition is satisfied only if the supersaturation activity ratio $\Omega$ is larger than a value $\idx{\Omega}{s}$ given by
\begin{equation}
    \label{eq:CNT_CriticalConfinedOmega_Steric}
    \idx{\Omega}{s} = \exp \left(
        \left[
            \frac{2 \sigmacl \vc}{\rp} - \DeltaPc \Delta v
        \right]
        /
        \left[
            \kb T
        \right]
    \right)
\end{equation}
The corresponding critical supersaturation $S = m / m_0$ can be found by inverting Eq. \ref{eq:SupersatActivity}, which requires numerical methods because both $S$ and $\gamma$ depend on the concentration.
A similar approach was developed in Ref. \cite{Jain2019}, however without evaluating the impact on kinetic nucleation as we do below.

For kinetic nucleation, using Eqs. \ref{eq:CNT_DeltaMuKinetics} and \ref{eq:DeltaMuConfined}, the critical supersaturation activity ratio is
\begin{equation}
    \label{eq:CNT_CriticalConfinedOmega_KineticRaw}
    \idx{\Omega}{k} = \exp \left(
        \mathcal{C} - \frac{\DeltaPc \Delta v}{\kb T}
    \right)
\end{equation}
While it is in principle possible to calculate $\mathcal{C}$ from physical parameters of the system (see Eq. \ref{eq:CNT_KineticConstant}), the task is not trivial due to significant uncertainties on the values for the interfacial energy $\sigmacl$ or the kinetic prefactor $K$ (see e.g. Ref. \cite{Cedeno2023}).
However, these parameters cancel out when considering ratios between the confined case and the bulk case under the same conditions (volume, time and kinetic prefactor).
With this approach, the combination of Eqs. \ref{eq:CNT_KineticsBulk} and \ref{eq:CNT_CriticalConfinedOmega_KineticRaw} directly leads to Eq. \ref{eq:CNT_Supersat} in the Main Text,
where we have expressed $\idx{\Omega}{\infty} = \left( \gamma_\infty S_\infty \right)^\nu$ using Eq. \ref{eq:SupersatActivity}.

\end{document}